\documentclass[lettersize,journal]{IEEEtran}
\usepackage{amsmath,amsfonts}
\usepackage{algorithmic}
\usepackage{array}
\usepackage[caption=false,font=normalsize,labelfont=sf,textfont=sf]{subfig}
\usepackage{textcomp}
\usepackage{stfloats}
\usepackage{booktabs}
\usepackage{float}
\usepackage{url}
\usepackage{verbatim}
\usepackage{graphicx}
\usepackage{color} % 使用color包
\usepackage[a4paper,left=1.8cm,right=1.8cm,top=2cm,bottom=4cm]{geometry}

\usepackage[colorlinks,
linkcolor=black,
anchorcolor=black,
citecolor=black,
urlcolor=blue]{hyperref}
\usepackage{balance}
\usepackage[numbers,sort&compress]{natbib}
\def\BibTeX{{\rm B\kern-.05em{\sc i\kern-.025em b}\kern-.08em
		T\kern-.1667em\lower.7ex\hbox{E}\kern-.125emX}}
\usepackage{balance}
\begin{document}
	
	%\makeatletter % 参考文献变色 <=======================================================
	%\let\myorg@bibitem\bibitem
	%\def\bibitem#1#2\par{%
		%  \@ifundefined{bibitem@#1}{%
			%    \myorg@bibitem{#1}#2\par
			%  }{%
			%    \begingroup
			%      \color{\csname bibitem@#1\endcsname}%
			%      \myorg@bibitem{#1}#2\par
			%    \endgroup
			%  }%
		%}
	%\newcommand*{\bibitem@sugiuraaaeffects}{blue}
	%\newcommand*{\bibitem@alamoutiaasimple}{blue}
	%\makeatother % 参考文献变色结束 <========================================================

	\title{Pre-Chirp-Domain Index Modulation for Affine Frequency Division Multiplexing}
	\author{Guangyao Liu$ ^1 $,
		Tianqi Mao$ ^{2,3} $,
		Ruiqi Liu$ ^4 $
		and Zhenyu Xiao$ ^1 $
		
		$^1$ Electronic and Information Engineering, Beihang University, \\Beijing 100191, China
		
		$^2$ MIIT Key Laboratory of Complex-Field Intelligent Sensing, Beijing Institute of Technology, \\Beijing 100081, China 
		
		$^3$ Yangtze Delta Region Academy, Beijing Institute of Technology (Jiaxing), \\Jiaxing 314019, China 
		
		$^4$ Wireless and Computing Research Institute, ZTE Corporation, Beijing 100029, China
		
		E-mails: $\{$liugy@buaa.edu.cn, maotq@bit.edu.cn, richie.leo@zte.com.cn, xiaozy@buaa.edu.cn $\}$
		
		\vspace{2mm}
	}
	
	%\markboth{Journal of \LaTeX\ Class Files,~Vol.~18, No.~9, September~2020}%
	%{How to Use the IEEEtran \LaTeX \ Templates}
	\pagestyle{empty}
	
	\maketitle
	\thispagestyle{empty}
	
	\begin{abstract}
		Affine frequency division multiplexing (AFDM), tailored as a novel multicarrier technique utilizing chirp signals for high-mobility communications, exhibits marked advantages compared to traditional orthogonal frequency division multiplexing (OFDM). AFDM is based on the discrete affine Fourier transform (DAFT) with two modifiable parameters of the chirp signals, termed as the pre-chirp parameter and post-chirp parameter, respectively. These parameters can be fine-tuned to avoid overlapping channel paths with different delays or Doppler shifts, leading to performance enhancement especially for doubly dispersive channel. In this paper, we propose a novel AFDM structure with the pre-chirp index modulation (PIM) philosophy (AFDM-PIM), which can embed additional information bits into the pre-chirp parameter design for both spectral and energy efficiency enhancement. Specifically, we first demonstrate that the application of distinct pre-chirp parameters to various subcarriers in the AFDM modulation process maintains the orthogonality among these subcarriers. Then, different pre-chirp parameters are flexibly assigned to each AFDM subcarrier according to the incoming bits. By such arrangement, aside from classical phase/amplitude modulation, extra binary bits can be implicitly conveyed by the indices of selected pre-chirping parameters realizations without additional energy consumption. At the receiver, both a maximum likelihood (ML) detector and a reduced-complexity ML-minimum mean square error (ML-MMSE) detector are employed to recover the information bits. It has been shown via simulations that the proposed AFDM-PIM exhibits superior bit error rate (BER) performance compared to classical AFDM, OFDM and IM-aided OFDM algorithms.

	\end{abstract}
	
	
	%\vfill
	\section{Introduction}
	\IEEEPARstart{T}{he} sixth-generation (6G) networks and beyond fifth-generation (B5G) wireless network systems are anticipated to facilitate reliable and high data rate communication for high-speed mobile scenarios, such as railway and Vehicle-to-Vehicle (V2V) communications \cite{liu20236g,liu2023beginning}. Under such mobility, wireless channels tend to experience doubly selective channel fading due to the coupling effects of Doppler shifts and multi-paths. These pose significant challenges to existing technologies, represented by OFDM that have been widely utilized in 4G/5G standards \cite{wu2016survey}. 
	
	Against this background,  a new multicarrier modulation scheme referred to as the affine frequency division multiplexing (AFDM) technique has been proposed \cite{bemani2021afdm}. AFDM is based on the discrete affine Fourier transform (DAFT), which is a generalized discrete Fourier transform. In AFDM, information symbols are multiplexed on a set of orthogonal chirps through DAFT and Inverse DAFT (IDAFT), which transforms the representation of the effective channel in the AFDM system from a linear time-varying (LTV) channel into a sparse, quasi-static channel.
	Therefore, AFDM can achieve complete diversity in dual dispersion channels \cite{bemani2023affine}. In the context of channel estimation, \cite{yin2022pilot} introduced a pair of pilot-assisted channel estimation methods. In \cite{bemani2022low}, two low complexity equalization schemes for AFDM under doubly dispersive channels were proposed. To the best of our knowledge, researches on spectrum and energy efficiency enhancement of the AFDM system are still at its infancy. 
	
	On the other hand, the index modulation (IM) has gained extensive interest due to its high spectral efficiency (SE) with low power consumption \cite{mao2018novel,ishikawa201850}. Distinct from traditional modulation schemes, IM can convey energy-free information bits through the activation patterns of various transmit entities, including the selection of subcarriers, the positioning of the pulse, and types of the APM constellation \cite{wen2019survey}. By such an arrangement, IM-assisted systems could achieve the same/higher throughput level with lower energy consumption than classical counterparts. Thanks to these merits, IM has found diverse applications in different 6G candidate technologies, such as terahertz (THz) communications \cite{mao2024index} and massive MIMO \cite{cui2016performance}.   
	
	In this paper, a novel AFDM structure with the pre-chirp index modulation (PIM) philosophy (AFDM-PIM) is proposed, which is capable of embedding additional information bits into the pre-chirp parameter design, thereby enhancing both spectral and energy efficiency. In particular, we first demonstrate that applying distinct pre-chirp parameters for different subcarriers in the AFDM modulation process preserves their orthogonality. Subsequently, we assign varying pre-chirp parameters to each AFDM subcarrier based on the incoming bits. This method allows for the transmission of extra binary bits through the indices of selected pre-chirping parameter realizations, without increasing energy consumption, in addition to the conventional phase/amplitude modulation. At the receiver, both a maximum likelihood (ML) detector and a reduced-complexity ML-minimum mean square error (ML-MMSE) detector are utilized for information bits recovery. Simulation results demonstrate that the proposed AFDM-PIM outperforms classical AFDM, OFDM, and IM-aided OFDM algorithms in terms of bit error rate (BER) performance.

	%The rest of this correspondence is organized as follows. In Section II, we present the input-output relationship of the AFDM scheme and explore the impact of AFDM parameters on the orthogonality of subcarriers. In Section III, we discuss both the proposed AFDM-PIM scheme and the detector. Our simulation results and comprehensive discussions are presented in Section IV, with the conclusion of this paper provided in Section V.

	%%%%%%%%%%%%%%%%%%%%%%%%%%%%%%%%%%%%%%%%%%%%%%%%%%%%%%%%%%%%%%%%%%
	%\vspace{10cm}
	%\newpage
	\section{System Model of AFDM}
	\subsection{DAFT-Based Modulation}
	AFDM is grounded in a generalized form of the discrete Fourier transform, known as the DAFT. After the bit stream enters the transmitter, it is mapped to $\mathbf x = [x[0],x[1],...,x[n]]$ in the discrete affine Fourier (DAF)-domain, where \( \mathbf x \in \mathbb{C}^{N \times 1} \) and \( N \) is the total number of subcarriers.
	
	After the mapping, the inverse DAFT (IDAFT) is performed on $\mathbf x$ to generate time-domain transmitted signals, which follows the three steps below \cite{healy2015linear}: 
	
	Initially, the pre-chirp multiplication is applied to the modulated symbols, yielding the frequency-domain signals $x'[m]$ as,
	\begin{equation}
		x'[m] = x[m] \cdot e^{i 2 \pi c_2 m^2}, \quad m=0, 1, \cdots, N-1,
	\end{equation}
	where $ c_2 $ is the pre-chirp parameter. This operation increases the dispersion of the signal on the time-frequency plane, which enhances the system performance under doubly dispersive channels.
	
	Subsequently, the IFFT is performed on the frequency-domain signals $x'[m]$, written as
	\begin{equation}
		s'[l] = \frac{1}{\sqrt{N}} \sum_{m=0}^{N-1} x'[m] \cdot e^{i 2 \pi \frac{ml}{N}}, \quad l=0, 1, \cdots, N-1,
	\end{equation}
	which transforms the chirp-processed frequency-domain signal to the time-domain samples $s'[l] (l=0,1,...,N-1)$.
	
	Finally, the post-chirp multiplication is is imposed on $s'[l]$, formulated as
	\begin{equation}
		s[l] = s'[l] \cdot e^{i 2 \pi c_1 n^2}, \quad l=0, \cdots, N-1,
	\end{equation}
	where $s[l]$ denotes time-domain transmitted signals, and $ c_1 $ is the post-chirp parameter. This operation further adjusts the time-frequency signal characteristics, bolstering the system's robustness against doubly dispersive channels. Based on (1-3), the transmitted AFDM signals $s[l]$ can be eventually derived as
	\begin{equation}
		s[l]=\frac{1}{\sqrt{N}}\sum_{m=0}^{N-1} x[m] \cdot e^{i 2 \pi\left(c_1 l^2+c_2 m^2+l m / N\right)}. \quad
	\end{equation}
	\subsection{Orthogonality Analysis of AFDM Subcarriers}
	From the modulation process of AFDM, we can rewrite (4) as
	\begin{equation}
		s[l]=\sum_{m=0}^{N-1}x[m]\Phi_l(m),\quad n=0, 1, \cdots,N-1,
	\end{equation}
	where $ \Phi_l(m) $ donates the subcarrier represented by
	\begin{equation}
		\Phi_l(m) = \frac{1}{\sqrt{N}} \cdot e^{i 2 \pi\left(c_1 l^2+c_2 m^2+l m / N\right)} 
	\end{equation}
	
	Thereby, for two subcarriers $\Phi_l^{c_1,c_{2,1}}(m)$ and $\Phi_l^{c_1,c_{2,2}}(m)$, which use the same $ c_1 $ but different value of $ c_2 $, we can derive the cross-correlation function as:
	\begin{equation}
		\begin{aligned}
			&\sum_{n=0}^{N-1}\Phi_l^{c_1,c_{2,1}}(m_1) \Phi_l^{c_1,c_{2,2}}(m_2) \\
			&=\frac1Ne^{-i2\pi\left(c_{2,1}m_1^2-c_{2,2}m_2^2\right)}\sum_{n=0}^{N-1}e^{-i\frac{2\pi}N(m_1-m_2)n} \\
			&=\frac1Ne^{-i2\pi\left(c_{2,1}m_1^2-c_{2,2}m_2^2\right)}\frac{1-e^{-i2\pi N\left(\frac{m_1-m_2}N\right)}}{1-e^{-i2\pi\left(\frac{m_1-m_2}N\right)}} \\
			&=0 (m_1 \neq m_2).
		\end{aligned}
	\end{equation}
	
	Based on the calculations discussed above, it is evident that employing different values of $ c_2 $ does not compromise the orthogonality among AFDM subcarriers. This insight lays a crucial groundwork for our forthcoming AFDM-PIM approach.
	%%%%%%%%%%%%%%%%%%%%%%%%%%%%%%%%%%%%%%%%%%%%%%%%%%%%%%%%%%%%%%%%%%
	%\newpage
	
	\begin{figure*}[!t]
		\centering
		\includegraphics[width=1\textwidth]{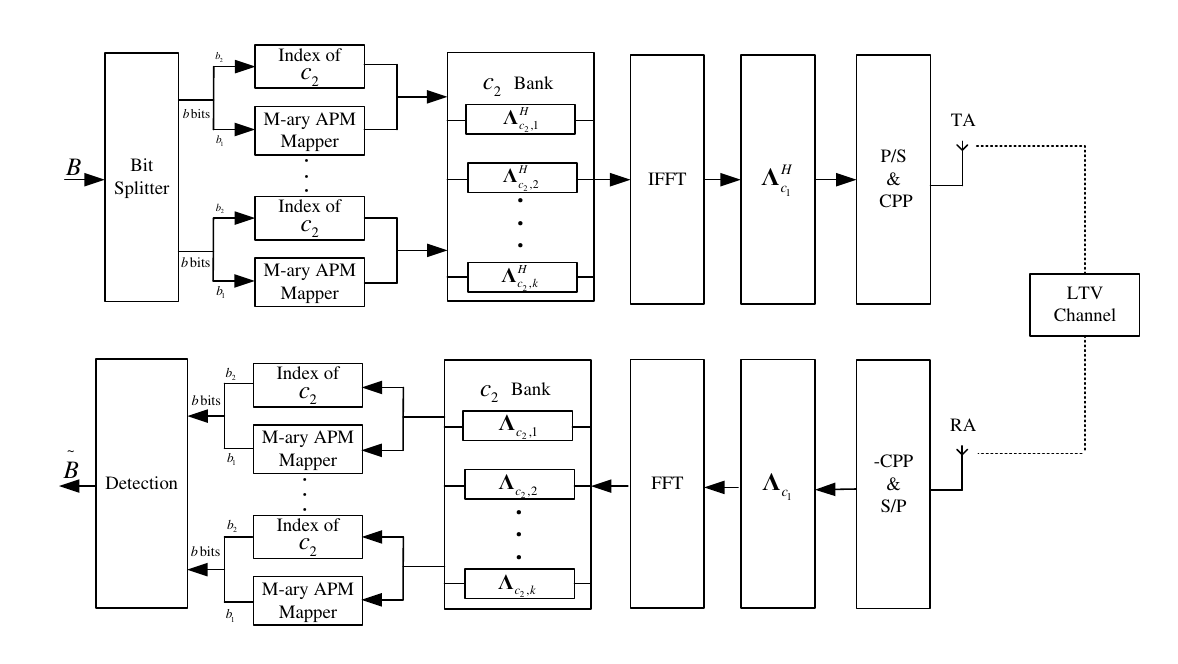}
		\caption{\quad Transceiver structure of the proposed AFDM-PIM scheme.}
		\label{System Model}
		
	\end{figure*}
	
	\section{Proposed AFDM-PIM Scheme}
	\subsection{Transmitter}
	The transceiver structure of the proposed AFDM-PIM scheme is presented in Fig. 1. Initially, for an AFDM-PIM system within $N$ subcarriers, the transmitted bits stream comprising $ B $ incoming bits is divided by a bit splitter into $ G $ subblocks. Each group consists of $ b $ bits and $n$ subcarriers, where \( b = \frac{B}{G} \) and $n=\frac{N}{G}$. Consider the $g$-th subblock $(g=0,1,...,G-1)$. The \( b \) information bits are partitioned into $b_1$ data bits and $b_2$ index bits. The data bits are modulated as $M_{mod}$-ary data symbols, denoted as $\mathbf{x}_g=[x_g(1), x_g(2), \cdots, x_g(n)] \in\mathbb{C}^{N\times1}$, which can be calculated as $ b_{1}=n\log_{2}(M_{mod}) $. On the other hand, each subcarrier selects a unique realization of $c_2$ from a pre-defined set $ \mathbb{S} =\left\{{S^{(1)}, S^{(2)}, \cdots, S^{Q_\mathbf{c_2}}} \right\}  $ according to the index bits. For clarity, we define the permutation of $c_2$ values for different subcarriers as \emph{pre-scaling pattern} (PSP). It can be seen that additional bits can be conveyed implicitly through indices of the chosen pre-chirping patterns, which is calculated as $ b_2=\left\lfloor\log_2(n!)\right\rfloor $. The $ \left\lfloor\ \right\rfloor $ represents the integer floor operator. Then the amount of transmitted bits for each group can be written as   
	\begin{equation}
		\begin{split}
			b &= b_{1} + b_{2}\\
			&=n\log_{2}(M_{mod}) + \left\lfloor\log_2(n!)\right\rfloor.
		\end{split}
	\end{equation}
	
	By denoting the PSP of the overall AFDM symbol as $\mathbf{c}_2=[c_{2,1}, c_{2,2}, ..., c_{2,m}] \in \mathbb{C}^{N\times1}$, the transmitted signals $s[l]$ for $l = 0,1,\ldots,N-1$ can be represented as the IDAFT of $\mathbf x$ using $\mathbf{c}_2$ for pre-chirping, formulated as
	\begin{equation}
		s[l]=\frac{1}{\sqrt{N}}\sum_{m=0}^{N-1} x[m] \cdot e^{i 2 \pi\left(c_1 	l^2+c_{2,m} m^2+l m / N\right)}, \quad
	\end{equation}
	where $ c_{2,m} $ represents the $c_2$ value for $m$-th subcarrier. In matrix form, the calculation of (9) can be expressed as
	\begin{equation}
		\mathbf{s}=\mathbf{D}^{\mathbf{H}}\mathbf{x}  = \boldsymbol{\Lambda}_{c_1}^H\mathbf{F}^H\boldsymbol{\Lambda}_{c_2}^H\mathbf{x},
	\end{equation}
	where $\boldsymbol\Lambda_{c_1}$ represents the post-chirp matrix,  $\mathbf{F}$ is the discrete Fourier transform (DFT) matrix with elements $\mathbf{F}(m,l) = e^{-i 2 \pi m l/N}/\sqrt{N}, m,l=0,1,\ldots, N-1$ and $\boldsymbol\Lambda_{c_2}$ represents the pre-chirp matrix. The $\boldsymbol\Lambda_{c_1}$ and $\boldsymbol\Lambda_{c_2}$ can be written as
	\begin{equation}
		\boldsymbol\Lambda_{c_1}=\operatorname{diag}\left(e^{-i 2\pi c_1l^2}, l=0,1,\ldots,N-1\right),
	\end{equation}
	\begin{equation}
		\boldsymbol\Lambda_{c_2}=\operatorname{diag}\left(e^{-i 2\pi c_{2,k}l^2},l,k=0,1,\ldots,N-1\right),
	\end{equation}
	respectively.
	
	Similarly to other multicarrier modulation technologies, AFDM-PIM also necessitates the inclusion of a prefix to effectively combat the effects of multi-path propagation. Taking advantage of the inherent periodicity characteristic of the DAFT, a chirp-periodic prefix (CPP) is employed in AFDM-PIM, which is expressed as
	\begin{equation}
		s[l]=s[N+l] e^{-i 2 \pi c_{1}\left(N^{2}+2 N l\right)},\quad l=-L_{\mathrm{cp}}, \cdots,-1,
	\end{equation}
	where the $ L_{\mathrm{cp}} $ is the length of CPP. Finally, the transmitted signal is forwarded to the receiver.
	\vspace{-2.6mm}
	\subsection{Receiver}
	Considering the time-frequency doubly dispersive channel, the received signals can be expressed as
	\begin{equation}
		r[l]=\sum_{p=1}^{P}s[l-p]h_l[d]+w[l],
	\end{equation}
	where $w[l]$ is the complex additive Gaussian noise (AWGN) and $h_l[d]$ represents the impulse response of the channel. The $h_l[d]$ can be modeled as 
	\begin{equation}
		h_l[d]=\sum_{p=1}^{P}h_{p}e^{-i 2\pi \nu_{p} l}\delta(d-d_{p}),		
	\end{equation}
	where $P$ represents the number of the multi-path and the $\delta(\cdot)$ denotes the Dirac delta function. Besides, $ \nu_{p} $, $ d_{p} $ and $h_{p}$ stand for the Doppler shift, integer time delay and the complex channel gain of the $p$-th path, respectively. $h_{p}$ follows a distribution of \(\mathcal{CN}(0,1/P)\). $d_p \in [0, d_{\max}] $, with \(d_{\max}\) being the maximum delay. The Doppler shift can be normalized as $ \alpha_{p} = N \Delta t \nu_{p}$, where $ \alpha_{p} \in [-\alpha_{\max},\alpha_{\max}]$, given that $ \alpha_{\max} $ and $\Delta t$ denote the maximum Doppler shift and sampling interval, respectively \cite{tao2023affine}.  For the integer Doppler shift value, the data bits are capable of achieving full diversity when the parameters follow that $c_1=\frac{2\alpha_{\max}+1}{2N}$ and $c_2$ is an irrational number \cite{bemani2023affine}.
	
	The matrix form of (14) is expressed as
	\begin{equation}
		\mathbf{r}=\mathbf{H}\mathbf{s}+\mathbf{w},
	\end{equation}
	where $\mathbf{w}=[w[0], w[1],\cdots,w[N-1]]\sim\mathcal{CN}(\mathbf{0},N_{0}\mathbf{I})$ denotes the $N\times 1$ noise vector and $\mathbf{H}$ stands for the time-domain channel matrix, calculated as $\mathbf{H}=\sum_{p=1}^{P}h_{p}\Gamma_{\mathrm{CPP}_{p}}\boldsymbol\Delta_{\nu_{p}}\mathbf\Pi^{d_{p}}$. $ \Gamma_{\mathrm{CPP}_{p}} $ is a diagonal matrix related to CPP. $\boldsymbol{\Delta_{\nu_{p}}}=\mathrm{diag}(e^{\large-j{2\pi}\nu_{p}l},l=0,1,\ldots,N-1)$ represents the Doppler shift of the $ p $-th path. $ \mathbf\Pi^{d_{p}} $ donates the the delay extension on the $ p $-th path.  
	
	After the removal of CPP, an N-point DAFT is performed on the $r$, yielding the received DAF-domain symbols formulated as
	\begin{equation}
		y[m]=\frac{1}{\sqrt{N}}\sum_{m=0}^{N-1} r[l] \cdot e^{-i 2 \pi\left(c_1 l^2+c_2^{(m)} m^2+l m / N\right)}, \quad
	\end{equation}
	which can be transformed into the matrix representation:
	\begin{equation}
		\mathbf{y}=\mathbf{D}\mathbf{r}=\sum_{p=1}^Ph_i\mathbf{D}\Gamma_{\mathrm{CPP}_p}\boldsymbol{\Delta}_{\nu_p}\mathbf{\Pi}^{d_p}\mathbf{D}^H\mathbf{x}+\mathbf{D}\mathbf{w}.
	\end{equation}
	
	The received signal from (18) can be further articulated using an effective channel matrix as
	\begin{equation}
		\begin{split}
			\mathbf{y} &= \mathbf{H}_{\mathrm{eff}}\mathbf{x}+\widetilde{\mathbf{w}}.\\
			&=\sum_{p=1}^P h_p \mathbf{H}_{p} \mathbf{x} + \widetilde{\mathbf{w}},
		\end{split}
	\end{equation}
	where $ \mathbf{H}_{\mathrm{eff}} = \mathbf{A}\mathbf{H}\mathbf{A}^H$ represents the effective channel matrix in DAF-domain,  $\mathbf{H}_{p}=\mathbf{\textsc{D}}\Gamma_{\mathrm{CPP}_{p}}\boldsymbol\Delta_{\nu_p}\mathbf\Pi^{d_{p}}\mathbf{D}^{H}$ represents the $ p $-th path channel and $\widetilde{\mathbf{w}}=\mathbf{A}\mathbf{w}$ is the effective noise. 
	
	Upon receiving the signal \(\mathbf{y} \), two detection methods are employed for data detection as follows:
	
	\textit{1) ML Detector:} The ML detector considers all possible signal realizations by searching for signal constellation points and the $c_2$ parameters indices.
	
	According to (19), the ML detector is given by
	\begin{equation}
		{\operatorname*{(\overset{\sim} {\mathbf{x}},\overset{\sim} {\mathbf {c_2}})}} = \underset{\mathrm{\forall \mathbf{x},\mathbf {c_2} }}{\operatorname*{\arg\min}}\left\|\mathbf{y-H_{eff}\mathbf{x}~}\right\|_F^2,
	\end{equation}
	The computational complexity of the ML detector in (20) is on the order of $\mathcal{O}(Q_\mathbf{c_2}M_{mod}^N)$ per frame, which increases exponentially with the number of AFDM subcarriers. To this end, we further develop a reduced-complexity detector.
	
	\textit{2) ML-MMSE Detector:} When the PSP of the AFDM-PIM is known, the received APM CS can be obtained by the MMSE detector as
	\begin{equation}
		\hat{\mathbf{x}}_{\mathrm{MMSE}}(k)=\mathbf{H}_{\mathrm{eff}}^{H}(k)(\mathbf{H}_{\mathrm{eff}}(k)\mathbf{H}_{\mathrm{eff}}^{H}(k)+\frac{1}{\gamma}\mathbf{I}_{N})^{-1}\mathbf{y},
	\end{equation}
	where $\gamma=E_{b}/N_{0}$ represents the average SNR and $k \in [1,Q_\mathbf{c_2}]$. Then,  identify the most likely received signal and PSP by comparing the outcomes of MMSE detector under various PSPs,
	\begin{equation}
		{\operatorname*{(\overset{\sim} {\mathbf{x}},\overset{\sim} {\mathbf {c_2}})}} = \underset{k \in [1,Q_\mathbf{c_2}]}{\operatorname*{\arg\min}}\left\|\mathbf{y-H_{eff}}(k)\hat{\mathbf{x}}_{\mathrm{MMSE}}(k)\right\|_F^2.
	\end{equation}
	The computational complexity of the ML-MMSE detector in (22) is on the order of $\mathcal{O}(Q_\mathbf{c_2}N^3)$ per frame.
	
	Finally, the bits stream can be reconstructed through the $c_2$ bank and the corresponding constellation demapper the estimation of PSP and the received APM CSs.
	\begin{figure}[!t]
		\flushleft 
		\includegraphics[width=3.6in]{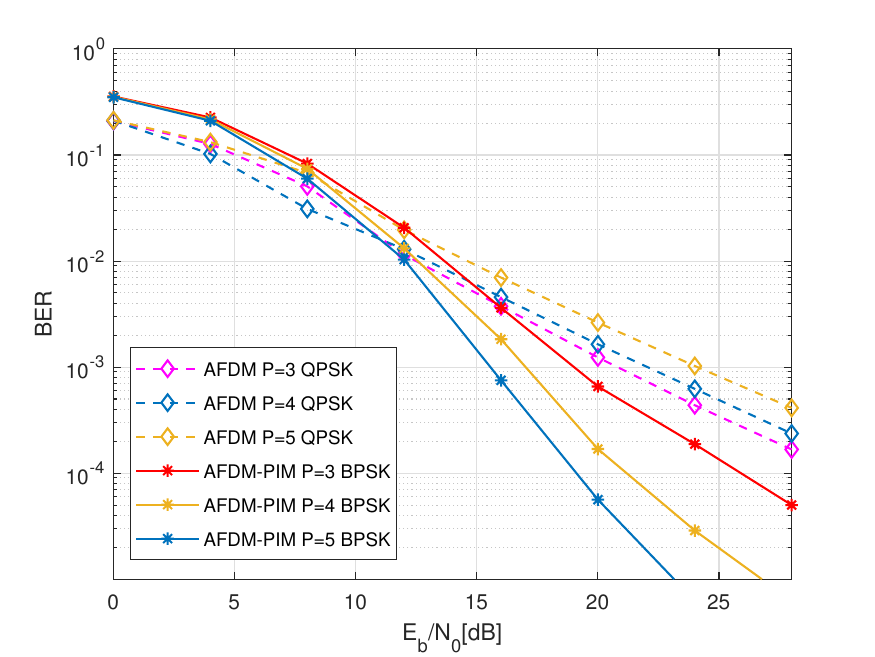}
		\caption{\quad BER performance comparison between the proposed AFDM-PIM and AFDM at 2 bits/s/Hz, where $N$=8, $ d_{\max} $ = 4 and $ \alpha_{\max} $ = 2.}
		\label{vsAFDM}
		\vspace{-3mm}
	\end{figure}
	%%%%%%%%%%%%%%%%%%%%%%%%%%%%%%%%%%%%%%%%%%%%%%%%%%%%%%%%%%%%%%%%%%
	%\newpage
	\section{Simulation Results and Analysis}
	In this section, we perform bit error rate (BER) simulation to assess the performance of the proposed AFDM-PIM scheme under various system configurations. For comparison, classical AFDM, OFDM and IM-aided OFDM schemes are also simulated as the benchmarks. In the simulation, the carrier frequency is 0.8 GHz and the subcarrier spacing is 1 KHz. Besides, we consider $c_1=\frac{2\alpha_{\max}+1}{2N}$ and we set the $\mathbf{c}_2$ in increments of $\frac{\pi}{2}$. For instance, $\mathbf{c}_2 = [\frac{\pi}{2}, \pi, \frac{3\pi}{2}, 2\pi]$ when $n=4$.
	
	Fig. 2 illustrates the comparison in the BER performance between the proposed AFDM-PIM and classic AFDM both with ML detection. The SE is maintained at 2 bits/s/Hz and we have the settings with $( N, g, n, d_{\max}, \alpha_{\max}) = (8, 2, 4, 4, 2)$. Then BPSK is employed in AFDM-PIM while AFDM utilizes QPSK. It can be observed that at the BER level of approximately $10^{-3} $, the AFDM-PIM scheme presents a performance gain of about 2 dB over its classical counterparts when P=3 is considered for multi-path effects. Besides, the performance enhancement of the proposed AFDM-PIM increases with the number of paths. This is because the AFDM-PIM scheme can achieve the same spectral efficiency by using a lower modulation order.
	\begin{figure}[!t]
		\centering
		\includegraphics[width=3.6in]{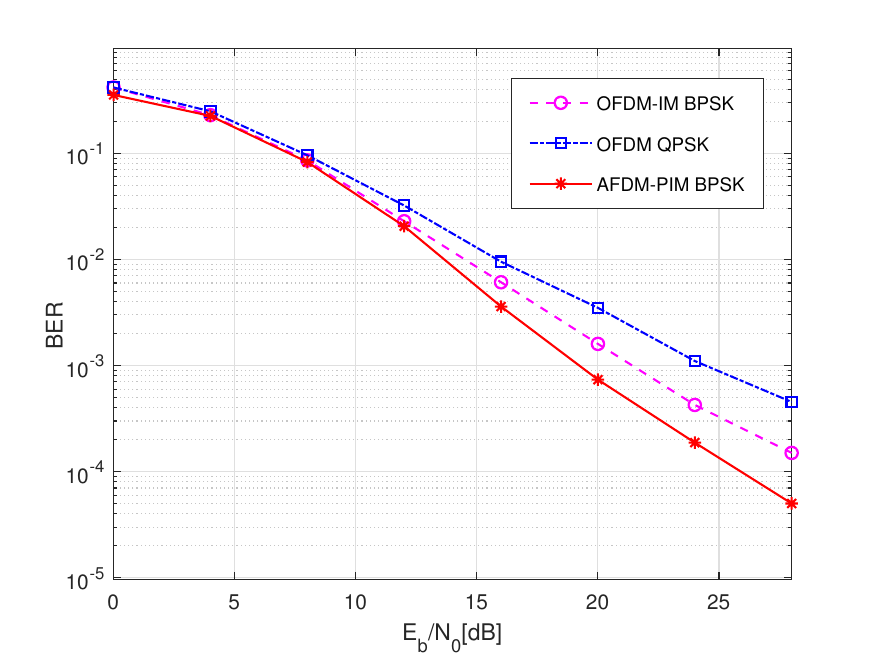}
		\caption{\quad BER performance comparison between the proposed AFDM-PIM and OFDM at 2 bits/s/Hz, where $N$=8, $P=3$, $ d_{\max} $ = 1 and $ \alpha_{\max} $ = 1.}
		\label{vsOFDM}
		
	\end{figure}
	
	In Fig.3, we investigate the BER performance of AFDM-PIM versus the classic OFDM and OFDM-IM in \cite{van2017impact}. For different schemes, the spectral efficiency is fixed as 2 bits/s/Hz, and ML detection is considered at the receiver. The simulation parameters are set as $( N, P, d_{\max}, \alpha_{\max}) = (8, 3, 1, 1)$. It can be observed from Fig. 3 that AFDM-PIM exhibits better performance by 5 dB than OFDM and 3 dB than OFDM-IM thanks to its capability of separating different paths in the DAF domain. This makes the proposed AFDM-PIM less the adverse effects of multipath propagation and Doppler shifts.
	\begin{figure}[!t]
		\centering
		\includegraphics[width=3.6in]{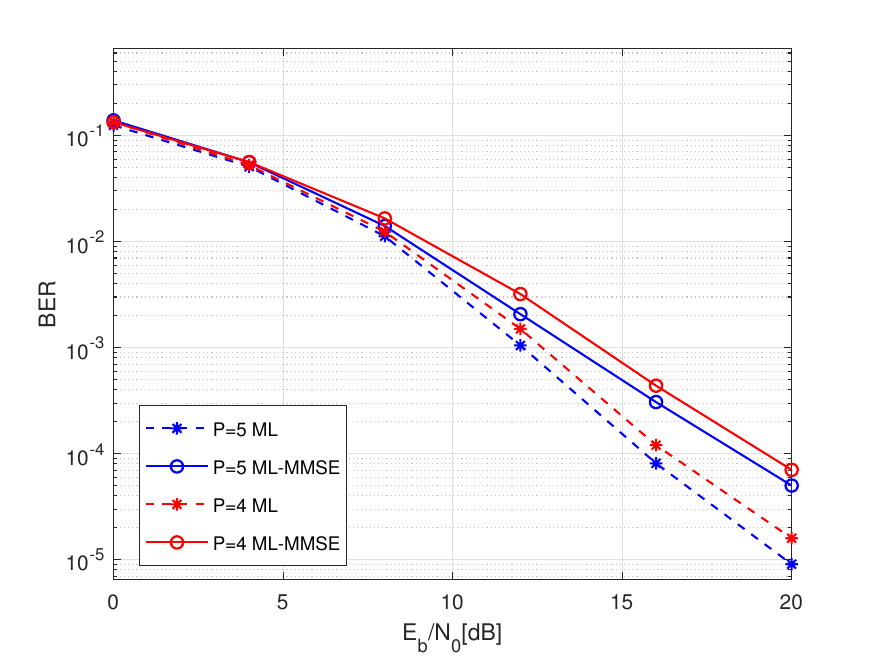}
		\caption{\quad BER performance comparison between the ML detector and ML-MMSE detector at 1.25 bits/s/Hz, where $N = 8 $, $ d_{\max} = 2 $, $ \alpha_{\max} = 2$ and BPSK is employed.}
		\label{vsIM-MMSE}
		
	\end{figure}
	
	Fig. 4 shows the BER performance comparison between the ML detector and the reduced-complexity ML-MMSE detector at a spectral efficiency of 1.25 bits/s/Hz. In this simulation, we have $N = 8 $, $ d_{\max} = 2 $, $ \alpha_{\max} = 2$ and $M_{mod} = 2$. The ML detector involves a global traversal of all possible PSP and data symbol realizations, offering theoretically optimal performance but at a significant increase in computational complexity. On the other hand, as shown in Fig. 4, the proposed ML-MMSE detector is capable of lowering the computational complexity with marginal performance loss of about 2 dB at the BER of $10^{-3} $, which validates its superiority for practical implementations.
	
	%%%%%%%%%%%%%%%%%%%%%%%%%%%%%%%%%%%%%%%%%%%%%%%%%%%%%%%%%%%%%%%%%%
	%\newpage
	\section{Conclusions}
	In this paper, we proposed a novel AFDM structure with the pre-chirp index modulation, which can leverage distinct pre-chirp parameters of AFDM to convey the index bits thus improving both spectral and energy efficiencies. The proposed scheme, which integrates the advantages of the DAFT in AFDM and the flexibility of index modulation, shows significant potential in high-mobility scenarios. Our comprehensive simulations demonstrate the superiority of AFDM-PIM over traditional OFDM and IM-aided OFDM schemes in terms of BER performance. The results highlight the scheme's robustness against the challenges posed by Doppler shifts and rapidly varying wireless channels. 
	
	\balance
	
	%Biography text here without a photo.
	
	%\begin{IEEEbiography}[{\includegraphics[width=1in,height=1.25in,clip,keepaspectratio]{fig1.png}}]{IEEE Publications Technology Team}
	%In this paragraph you can place your educational, professional background and research, and other interests.\end{IEEEbiography}
\footnotesize
\bibliographystyle{IEEEtran}   %指定参考文献样式文件为IEEEtran.bst
\bibliography{AFDM-PIM}   %指定所使用的bib文件

\end{document}